\documentclass[acmsmall,screen,nonacm]{acmart}
\setcopyright{none}
\authorsaddresses{}

\AtBeginDocument{%
  }

\setcopyright{acmlicensed}
\copyrightyear{2018}
\acmYear{2018}
\acmDOI{XXXXXXX.XXXXXXX}
\acmConference[Conference acronym 'XX]{Make sure to enter the correct
  conference title from your rights confirmation email}{June 03--05,
  2018}{Woodstock, NY}
\acmISBN{978-1-4503-XXXX-X/2018/06}

\usepackage{changepage}
\usepackage{framed}
\usepackage{tabularx}
\usepackage{subcaption}
\usepackage{multirow} 
\usepackage{tabularx}
\usepackage[table]{xcolor}
\usepackage{lipsum} 
\usepackage[acronym,nohypertypes={acronym}]{glossaries}
\makeatletter
 {\par\unskip\endMakeFramed}
\makeatother

\definecolor{formalshade}{rgb}{0.93,0.93,0.93}
\definecolor{darkblue}{rgb}{0.2, 0.2, 0.2}

\newenvironment{formal}{%
  \def\FrameCommand{%
    \hspace{1pt}%
    {\color{darkblue}\vrule width 2pt}%
    {\color{formalshade}\vrule width 4pt}%
    \colorbox{formalshade}%
  }%
  \MakeFramed{\advance\hsize-\width\FrameRestore}%
  \noindent\hspace{-1pt}% disable indenting first paragraph
  \begin{adjustwidth}{}{7pt}%
  \vspace{2pt}\vspace{2pt}%
}
{%
  \vspace{3pt}\end{adjustwidth}\endMakeFramed%
}

\newcounter{resultcounter}
\newenvironment{result}{\begin{formal}
    \refstepcounter{resultcounter}
    \noindent \hspace{-4pt}
}{
\end{formal}
}
\usepackage{hyperref,url}
\usepackage{xspace}
\newcommand{\etal}[1]{#1~\textit{et al.}}
\newcommand\ie{\emph{i.e.},\xspace}

\usepackage{pifont}% http://ctan.org/pkg/pifont

\usepackage{booktabs}
\usepackage{makecell}

\usepackage{flushend}
\newacronym{spl}{SPL}{{\emph{Software Product Line}}}
\newacronym{ttft}{TTFT}{\emph{Time to First Token}}
\newacronym{itl}{ITL}{\emph{Inter-Token Latency}}
\newacronym{vllm}{vLLM}{\emph{virtual Large Language Model}}
\newacronym{tgi}{TGI}{\emph{Text Generation Inference}}
\newacronym{moe}{MOE}{\emph{Mixture of Experts}}
\newacronym{ast}{AST}{\emph{Abstract Syntax Trees}}
\newacronym{msr}{MSR}{\emph{Mining Software Repositories}}
\newacronym[
    \glslongpluralkey={\emph{Large Language Models}}, 
    \glsshortpluralkey={LLMs}
]{llm}{LLM}{\emph{Large Language Model}}

\begin{document}

%\title{What's in a Configuration: An Empirical Study of LLM Inference Configuration Practices}
\title{How Developers Configure LLM Inference: A Large-Scale Study of Real-World Practices}

\author{Nada Zine}
\email{nada.zine@inria.fr}
\affiliation{%
  \institution{Univ. Lille, CNRS, Inria, UMR 9189 CRIStAL}
  \city{Lille}
  \country{France}
}

\author{Matias Martinez}
\email{matias.martinez@upc.edu}
\affiliation{%
  \institution{Universitat Politècnica de Catalunya}
  \city{Barcelona}
  \country{Spain}
}

\author{Romain Rouvoy}
\email{romain.rouvoy@inria.fr}
\affiliation{%
  \institution{Univ. Lille, CNRS, Inria, UMR 9189 CRIStAL}
  \city{Lille}
  \country{France}
}

\author{Clément Quinton}
\email{clement.quinton@inria.fr}
\affiliation{%
  \institution{Univ. Lille, CNRS, Inria, UMR 9189 CRIStAL and IUF}
  \city{Lille}
  \country{France}
}
\begin{abstract}

\glspl{llm} are increasingly integrated into software systems, making deployment and inference configuration a recurring software engineering task. 
Inference backends expose many options that affect model loading, generation, serving, and request processing. 
Yet little is known about which of these options developers explicitly configure in practice.
In this paper, we present a large-scale empirical study of \gls{llm} inference configuration practices in open-source software using Hugging Face Transformers, \gls{vllm}, and the OpenAI API. 
We analyze $5,398$ GitHub repositories using at least one of these backends and identify $345,117$ explicitly configured option occurrences.
Our results show that developers use only a small part of the available configuration space regularly. Although $98.8\%$ of the documented backend-specific options appear at least once, $78.9\%$ are configured in at most $10\%$ of repositories using the corresponding backend. Configurations are also small, with a median of $2$ options for Transformers and $4$ for \gls{vllm} and OpenAI, but the combinations of options vary widely: $56.8$--$61.6\%$ of multi-option signatures appear in only one repository. We also observe recurring option values as well as substantial variation in how values are chosen. Finally, configuration practices differ across application contexts, with option-adoption differences reaching $30.7$ percentage points.
These findings characterize the configuration surface developers make explicit rather than the importance or optimality of individual options. They provide empirical evidence for prioritizing future studies of inference configuration and for improving backend documentation and configuration support.
\end{abstract}



\maketitle
\glsresetall
\section{Introduction}
\label{sec:introduction}

\glspl{llm} such as GPT-series~\cite{achiam2023gpt}, Qwen~\cite{yang2025qwen3}, and Mistral~\cite{jiang2024mixtral} are increasingly integrated into modern software systems across applications such as coding assistants, content generation, search, and healthcare~\cite{bharathi2024analysis,murtaza2025impact}.
Integrating these models requires running \gls{llm} inference, during which a trained model processes an input prompt or context to generate an output, typically token by token.
As inference becomes a recurring software engineering task, developers must make configuration choices that can directly affect latency, throughput, energy consumption, monetary cost, and output quality~\cite{jegham2025hungry,shekhar2024towards}.
These choices are particularly important because inference accounts for over $90\%$ of overall \gls{llm} compute cycles~\cite{patel2024characterizing}.
Developers must therefore carefully configure inference to balance application efficiency and behavior.

These choices are complicated by the large configuration spaces of modern inference backends.
For example, Hugging Face Transformers~\cite{wolf2019huggingface}, \gls{vllm}~\cite{kwon2023efficient}, Ollama~\cite{ollama}, llama.cpp~\cite{llamacpp}, and the OpenAI API\footnote{\url{https://developers.openai.com/api/reference/python}} expose dozens to hundreds of options spanning model loading, decoding, resource management, serving, and request processing~\cite{park2025survey}.
Developers must decide which options to configure, which defaults to retain, which values to select, and which options to combine, while the available choices also differ across backends.

This large configuration space is difficult to study exhaustively. Evaluating all options, values, and their interactions is infeasible because of the combinatorial number of possible configurations and the cost of running such experiments. Existing empirical studies therefore focus on selected subsets of configuration options and values~\cite{zine2026pimp}.

\emph{However, these studies tell us little about which parts of the configuration space developers actually use in real software.}
Options that strongly affect performance in controlled experiments may rarely be configured in practice, while commonly configured options or combinations may receive comparatively little attention.
Despite the growing body of work on \gls{llm} inference, we therefore lack a systematic understanding of which options developers configure, which values they select, how they combine options, and how these practices vary across backends and application contexts.
\emph{Without this evidence, it is difficult to assess how well inference research reflects developer practice and which parts of the configuration space deserve further investigation or better tooling support.}

In this paper, we address this gap by studying how \gls{llm} inference is configured in open-source software.
We study three widely used inference backends: Hugging Face Transformers, \gls{vllm}, and the OpenAI API.
We analyze $5,398$ GitHub repositories using at least one of these backends and identify $345,117$ explicitly configured option occurrences.
We study what developers configure, how they compose configurations, and how configuration practices vary across application contexts.

Our results show a clear difference between the options available to developers and those they commonly configure. While $98.8\%$ of the documented backend-specific options appear at least once in our dataset, $78.9\%$ are configured in at most $10\%$ of repositories using the corresponding backend. Developers generally create small configurations, with a median of $2$ options for Transformers and $4$ for \gls{vllm} and OpenAI, but the combinations they use vary widely: $56.8$--$61.6\%$ of multi-option signatures occur in only one repository. We also observe recurring option values alongside substantial variation in value choices, as well as differences across application contexts of up to $30.7$ percentage points in option adoption. Together, these results provide a first large-scale view of real-world LLM inference configuration practices and identify concrete opportunities for future studies, tooling, and backend support.

The main contributions of this paper are:

(i) \textbf{Empirical characterization.}
We provide the first large-scale characterization of \gls{llm} inference configuration practices across $5,398$ open-source Github repositories, covering option adoption, value distributions, recurring combinations, and application contexts;

(ii) \textbf{Configuration patterns.}
We show that most available options are rarely configured, while individual configurations remain small but their option combinations are highly diverse.
We further identify backend- and context-specific configuration patterns;

(iii) \textbf{Mining pipeline and dataset.}
We develop a static-analysis pipeline for extracting explicitly configured inference options, values, and combinations from thousands of projects, and release the resulting dataset and replication package.

Our study targets software developers, inference backend maintainers, and researchers seeking to understand real-world configuration practices, improve configuration support, and prioritize future studies of \gls{llm} inference.
\section{Related Work}\label{sec:related-work}

\paragraph{\textbf{Configurable LLM Inference Backends}}
Recent \gls{llm} inference backends expose configuration choices spanning model loading, memory management, batching, parallelism, caching, quantization, and generation. 
These interfaces build on optimization techniques such as continuous batching~\cite{yu2022orca}, PagedAttention~\cite{kwon2023efficient}, FlashAttention~\cite{dao2022flashattention,dao2023flashattention,ye2025flashinfer}, prefix caching~\cite{zheng2024sglang}, chunked prefill~\cite{agrawal2024taming}, prefill--decode separation~\cite{zhong2024distserve}, and quantization~\cite{frantar2022gptq}. 
Depending on the backend, these choices appear as API parameters, command-line arguments, configuration files, or environment variables. 
Consequently, inference optimization techniques create a large and evolving configuration space.

\textbf{Experimental Studies of LLM Inference Configurations}
Prior work has extensively evaluated how inference choices affect performance, efficiency, and output quality.
Infrastructure-level studies have examined GPU characteristics, power-management settings, model parallelism, batching, concurrency, placement, and cluster architecture~\cite{patel2024characterizing,stojkovic2024towards,stojkovic2025dynamollm,lazuka2024llm,martinez2025impact}. 
Model and workload-level studies have investigated model size, quantization, serving frameworks, prompt and generation lengths, request distributions, and task characteristics~\cite{samsi2023words,luccioni2024power,maliakel2025investigating,wilkins2024offline,husom2025sustainable,delavande2026understanding,fu2025llmco2,zine2026attention}. 
Other studies have focused on decoding and generation parameters such as temperature, top-k, top-p, beam search, and speculative decoding~\cite{shi2402thorough,arias2025decoding,nik2025impact,fernandez2025energy,zine2026pimp}. 
Overall, these studies show that inference behavior depends on interacting choices across models, workloads, hardware, and optimization objectives. However, they primarily evaluate configurations selected \textit{a priori} by researchers or optimization systems.
They therefore address which configuration choices affect inference outcomes, whereas we investigate which parts of the available configuration space developers explicitly configure in real-world application software.

\textbf{Mining Configuration Practices in Software and Machine Learning} 
Researchers have used repository mining and static analysis to study configuration in software systems from several perspectives.
\etal{Nadi} extracted constraints among configuration options~\cite{nadi2014mining}, \etal{Sayagh} studied how configuration options and frameworks are engineered and maintained~\cite{sayagh2017does,sayagh2018software}, and \etal{Bessghaier} showed that deployment and runtime configurations can span heterogeneous software artifacts~\cite{bessghaier2023constitutes}.
Related work has mined machine-learning repositories to characterize development practices and architectures~\cite{gonzalez2020state,nahar2025product}. 

Most closely related to our study, \etal{Simon} statically analyzed $2,000$ machine-learning repositories and associated papers and found that developers explicitly set only a subset of available hyperparameters and frequently rely on default values~\cite{simon2023exploring}. 
While their work focuses on model hyperparameters and their realization in ML implementations, we study inference-time configuration in application software, including option adoption, configured values, multi-option signatures, differences among inference backends, and associations with application contexts.

To the best of our knowledge, prior work has not systematically mined a large corpus of application repositories to characterize how developers explicitly use the configuration interfaces exposed by \gls{llm} inference backends. 
We complement experimental inference studies by providing evidence about which options, values, and combinations occur in practice, and complement configuration-mining studies by focusing specifically on \gls{llm} inference configuration.
\section{Study design}\label{sec:methodology}
We designed our study in accordance with established empirical software engineering practices~\cite{shull2007guide} and prior work on mining software repositories~\cite{tutko2022software}. 
In particular, we follow a structured study design based on explicit research questions, predefined repository-selection criteria, data extraction and validation procedures, and a reproducible analysis pipeline. 
To support independent verification and replicability, we provide a complete replication package~\cite{replicationpackage} containing all paper-related data, including the source code of our mining and analysis pipeline, the collected repository metadata, the extracted inference configurations, and the scripts used to produce our results.

\subsection{Terminology}
\begin{figure*}[t]
    \centering

    \begin{minipage}[t]{0.59\textwidth}
        \centering
        \includegraphics[width=\linewidth]{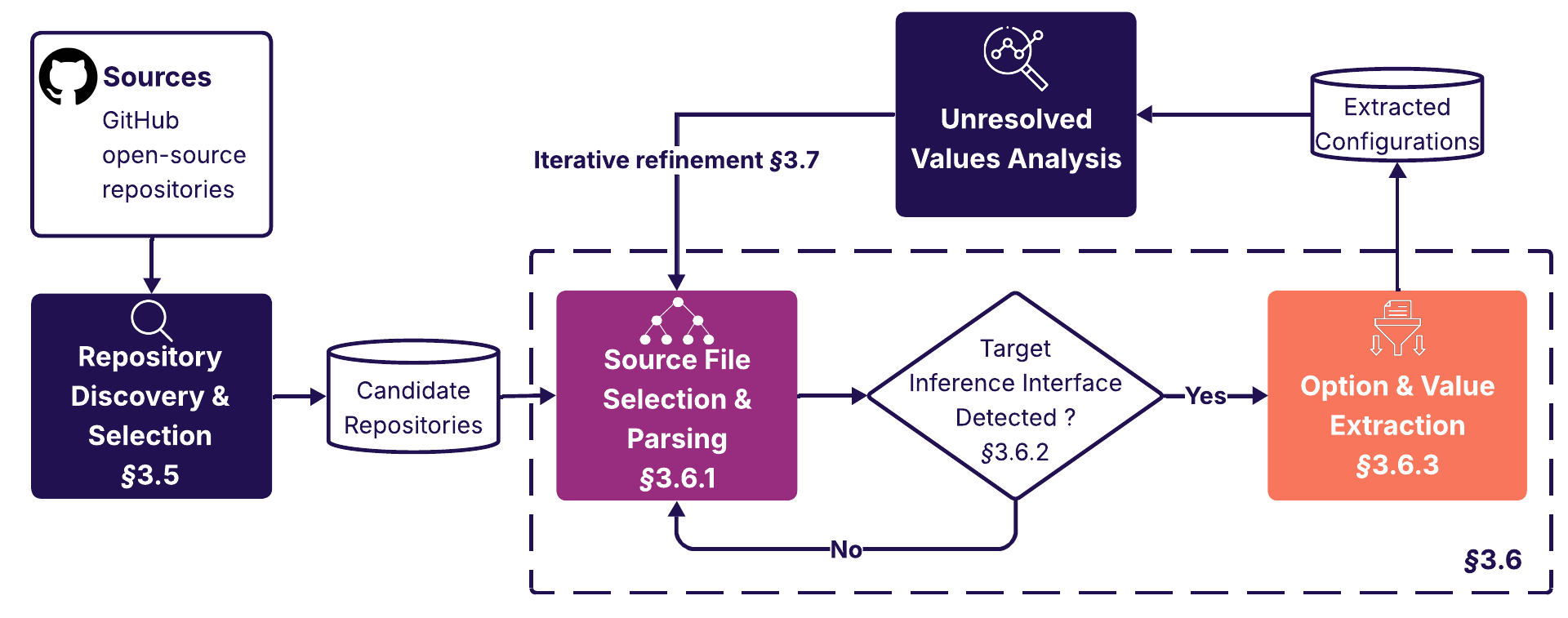}
        \captionof{figure}{Overview of the repository mining and configuration extraction pipeline.}
        \label{fig:overview}
    \end{minipage}
    \hfill
    \begin{minipage}[t]{0.39\textwidth}
        \centering
        \includegraphics[width=\linewidth]{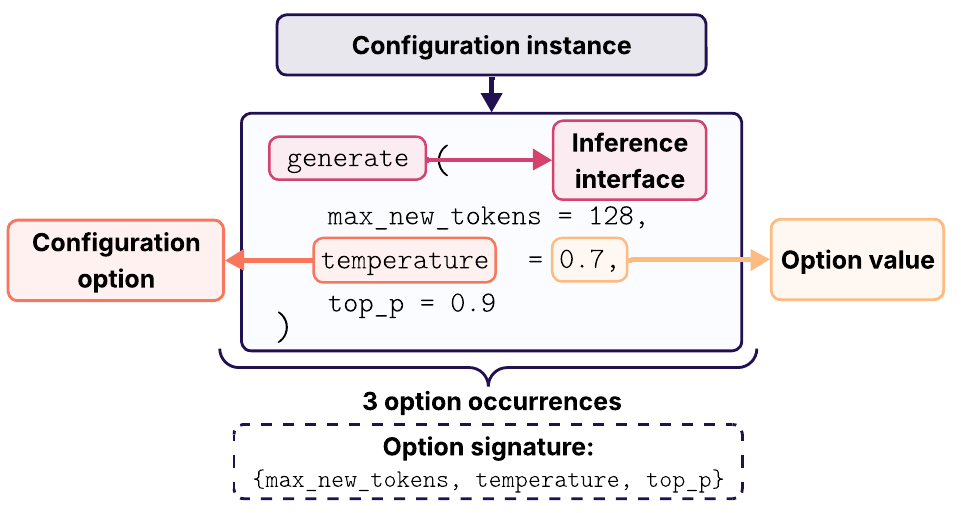}
        \captionof{figure}{Illustration of the terminology used throughout the paper.}
        \label{fig:terminology}
    \end{minipage}

\end{figure*}

We use \emph{inference backend} to refer to a software system or API through which LLM inference is performed. 
As illustrated in Figure~\ref{fig:terminology}, an \emph{inference interface} is a backend entry point through which inference is invoked or configured, while a \emph{configuration instance} corresponds to one concrete use of such an interface. 
Each explicitly specified \emph{configuration option} and its associated \emph{option value} form an \emph{option occurrence}. 
The \emph{option signature} of a configuration instance is the set of option names it explicitly specifies, regardless of their values. 
A \emph{singleton signature} denotes a signature used by only one repository.

\subsection{Goal \& Research Questions}
This study analyzes \gls{llm} inference configurations in real-world software projects to understand how developers configure \gls{llm} inference in practice and to provide insights for software developers, inference backend maintainers, and software engineering researchers studying open-source software.
More broadly, our study is guided by the following research question: \emph{How is the large configuration space exposed by \gls{llm} inference backends actually used by software developers?}
We investigate this question along three complementary dimensions, reflected in the following research questions.

\paragraph{\textbf{RQ1: How do developers configure \gls{llm} inference options in practice?}}
We investigate this question through two complementary dimensions: which options developers explicitly configure and which values they assign.

\paragraph{RQ1.1: Which \gls{llm} inference options do developers explicitly configure in practice?}
Despite the broad configuration space exposed by modern inference backends, little is known about which options developers explicitly configure and whether these choices differ across backends.

\paragraph{RQ1.2: Which values do developers assign to these inference options?}
Beyond identifying which options are explicitly configured, examining the values developers choose helps determine whether configuration practices converge around a small set of commonly used values or remain diverse.

\paragraph{\textbf{RQ2: How are real-world inference configurations structured in terms of the number of configured options and their combinations?}}
Analyzing individual options alone does not capture how developers construct complete inference configurations, where several options may be specified together. 
This research question examines how inference configurations are structured and whether recurring option combinations emerge across inference backends.

\paragraph{\textbf{RQ3: How do configuration practices differ across application contexts?}}
Configuration choices may vary depending on how \glspl{llm} are used within software projects. 
This research question is exploratory: rather than constructing an exhaustive taxonomy of LLM applications, we use available GitHub topics to examine whether explicit configuration practices are broadly shared across projects or associated with particular application contexts.

\subsection{Pipeline overview}
Figure~\ref{fig:overview} summarizes the main steps of our mining and analysis pipeline.
Starting from GitHub open-source repositories, we first identify candidate projects through API-based search and apply selection criteria to build our repository dataset.
We then analyze each repository's source code by parsing it, identifying inference interfaces, mapping them to their configuration options, and extracting the corresponding configurations and values.
We transform the extracted configurations into structured data for the analyses used to answer our research questions.
The pipeline includes a refinement loop in which we manually examine recurring unresolved values to improve backend-specific interface mappings and extraction rules.
The following subsections describe these steps in more detail: we first define the inference backends and configuration options covered by the study, then explain how repositories are discovered and selected, how configurations are extracted from source code, and how we analyze the resulting data with respect to our research questions.

\subsection{Inference Configuration Scope}\label{subsec:inference-configuration-scope}
\subsubsection{Selected Inference Backends}
For our study, we focus on three widely used~\cite{jiang2025foundation,chen2026developers} \gls{llm} inference backends: Hugging Face Transformers library~\cite{wolf2019huggingface}, \gls{vllm}~\cite{kwon2023efficient}, and OpenAI Python SDK~\cite{openai-python}. 
We select these backends because they capture complementary ways of performing \gls{llm} inference, respectively covering local model inference, high-throughput model serving, and API-based inference.
Hugging Face Transformers is a widely adopted open-source library that provides pretrained and fine-tuned models for a broad range of machine learning tasks and supports both training and inference. 
It can automatically download model weights and configurations and load them for local inference.
\gls{vllm} is an open-source \gls{llm} inference and serving engine designed for high-throughput and memory-efficient model execution. 
It can load models from the Hugging Face ecosystem and provides optimized capabilities for offline inference and online serving, making it a state-of-the-art solution for \gls{llm} inference and serving~\cite{su2025seesaw}. 
Finally, we include the OpenAI Python SDK to capture API-based inference practices that are common in application-level \gls{llm} software. 
Beyond accessing OpenAI-hosted models, the SDK can be configured to interact with other cloud-based or locally deployed providers that expose an OpenAI-compatible API, including serving engines such as \gls{vllm}.

\subsubsection{Selected Interfaces and Configuration Options}\label{subsubsec:selected-options}
This subsection defines which configuration options we recognize as inference-related after analyzing the source code. 
We define them from the documented inference-related interfaces of each selected backend. We rely on the documented configuration spaces exposed by Transformers version 5.12.0, \gls{vllm} version 0.23.0, and the OpenAI Python SDK version 2.41.1. 
These versions correspond to the latest stable releases available at the time of our study.
For each backend, we reviewed its official documentation\footnote{
Transformers: \url{https://huggingface.co/docs/transformers/};
vLLM: \url{https://docs.vllm.ai/};
OpenAI API: \url{https://platform.openai.com/docs/api-reference/}.
}
to identify the public classes, functions, and endpoints explicitly intended for model loading, inference, generation, or serving. 
We then collected the documented parameters exposed by these interfaces, which define the inference configuration options studied in this work.

For Hugging Face Transformers, we extracted parameters associated with the main inference-related classes and functions, including \texttt{GenerationConfig}, \texttt{pipeline}, \texttt{AutoModelForCausalLM}, \texttt{AutoTokenizer}, and \texttt{AutoModel}.
For \gls{vllm}, we focused on the primary inference and serving interfaces, namely \texttt{SamplingParams}, \texttt{LLM}, \texttt{EngineArgs}, and \texttt{AsyncEngineArgs}.
For the OpenAI API, we extracted parameters from the three text-generation endpoints exposed by the OpenAI Python SDK: \texttt{responses.create}, \texttt{chat.completions.create}, and \texttt{completions.create}. We considered the union of parameters accepted by these endpoints to cover both the current recommended endpoint, \ie Responses API, and the older Chat Completions and Completions interfaces, which may still be used in open-source repositories.

Table~\ref{tab:selected-parameter-groups} summarizes the main categories of configuration options covered by our study across Hugging Face Transformers, \gls{vllm}, and the OpenAI API. 
To construct this classification, we first relied on the categories provided by the Hugging Face Transformers documentation and on the taxonomy proposed by \etal{Park}~\cite{park2025survey}. 
Because these existing classifications cover only part of the broader configuration space considered in our study, we adapted and extended them to account for additional concerns exposed by the selected interfaces. 
We assigned each option to a category based on its documented primary purpose. 
The second author independently reviewed the resulting assignments, and we discussed potentially ambiguous cases until we reached agreement. 
The complete classification, including each option's assigned category, provenance, and the rationale for each assignment, is provided in our replication package. 
The package also reports which backends expose at least one option from each category.

\begin{table}[t]
\centering
\scriptsize
\caption{Inference option categories and representative options.}
\label{tab:selected-parameter-groups}

\begin{tabularx}{\columnwidth}{X X}
\toprule
\textbf{Category} & \textbf{Representative options} \\
\midrule

Model/loading &
\texttt{model}, \texttt{dtype}, \texttt{device\_map} \\

Compression &
\texttt{quantization}, \texttt{kv\_cache\_dtype} \\

Length control &
\texttt{max\_tokens}, \texttt{max\_output\_tokens} \\

Decoding/sampling &
\texttt{temperature}, \texttt{top\_p}, \texttt{top\_k} \\

Logit manipulation/constraints &
\texttt{repetition\_penalty}, \texttt{frequency\_penalty} \\

Stopping/special tokens &
\texttt{stop}, \texttt{stop\_token\_ids} \\

Serving/performance &
\texttt{tensor\_parallel\_size}, \texttt{gpu\_memory\_utilization} \\

Input/request &
\texttt{prompt}, \texttt{messages}, \texttt{input} \\

Structured output/tools &
\texttt{response\_format}, \texttt{tools} \\

Reasoning &
\texttt{reasoning}, \texttt{reasoning\_effort} \\

Output/response &
\texttt{logprobs}, \texttt{output\_scores} \\

Multimodal/media &
\texttt{audio}, \texttt{modalities} \\

Adapters/LoRA &
\texttt{enable\_lora}, \texttt{max\_loras} \\

API/request controls &
\texttt{metadata}, \texttt{timeout} \\

Safety/moderation &
\texttt{safety\_identifier}, \texttt{moderation} \\

Monitoring/profiling &
\texttt{profiler\_config}, \texttt{kv\_cache\_metrics} \\

\bottomrule
\end{tabularx}
\end{table}

\subsection{\textbf{Repository Discovery \& Selection}}\label{subsec:repository-collection-selection}
\subsubsection{GitHub API Search Strategy}
We collected repositories from GitHub using the GitHub Search API.
We ran the final mining run in mid-September 2026, providing a snapshot of the GitHub ecosystem at that time.
Based on the inference backends defined above, we designed backend-specific search queries for Hugging Face Transformers, \gls{vllm}, and OpenAI to identify candidate repositories.
For Hugging Face Transformers, the queries include terms such as \texttt{transformers}, \texttt{huggingface transformers}, \texttt{text generation}, \texttt{llm inference}, and \texttt{causal lm}.
For \gls{vllm}, they include terms related to inference serving, such as \texttt{vllm}, \texttt{llm serving}, \texttt{openai-compatible}, \texttt{vllm server}, and \texttt{inference server}.
OpenAI includes both general terms, such as \texttt{openai} and \texttt{openai api}, and API-specific expressions, such as \texttt{chat.completions} and \texttt{responses.create}, \texttt{from openai import OpenAI}, \texttt{client.chat.completions.create}, and \texttt{openai.ChatCompletion.create}.

\subsubsection{Candidate Repository Selection}
We selected candidate repositories according to several criteria. 
We considered only publicly accessible GitHub repositories to support reproducibility~\cite{kalliamvakou2014promises}. 
The analysis was restricted to Python repositories, thereby focusing on Python-based \gls{llm} inference ecosystems. 
We chose Python because all three studied inference backends provide Python interfaces and because Python is the predominant language in open-source AI and machine-learning software~\cite{gonzalez2020state}. 
To retain actively maintained and relevant projects, we required recent development activity, defined as activity on or after January $1$, $2025$. 
We also excluded archived and forked repositories and kept only repositories with at least $50$ stars.
Finally, we discarded repositories whose names contained terms such as \texttt{tutorial}, \texttt{demo}, or \texttt{example}, since these are more likely to correspond to pedagogical or demonstrative artifacts than to actual usage projects. 

At the end of this stage, the selected repositories form a filtered candidate dataset. 
They are not yet considered confirmed \gls{llm} inference repositories, as GitHub API Search may return false positives. 
For example, vLLM may simply be mentioned in a documentation file without actually being used in the source code.

\subsection{\textbf{Configuration Extraction}}\label{subsec:static-analysis-configuration-extraction}
After candidate selection, we analyze each repository's source code to confirm the use of the selected inference backends and extract their inference configurations.

\subsubsection{Source File Selection \& Parsing}
For each candidate repository, we retrieve its contents and inspect its Python source files. 
We retain files with a Python extension (\texttt{.py}) and exclude paths unlikely to contain application-level inference code, such as tests and documentation. 
Each retained source file is parsed into an \gls{ast} using Python's standard \texttt{ast} library. 
Files that cannot be parsed are skipped and recorded as analysis errors.
The resulting \glspl{ast} constitute the source representation used by the subsequent static analysis.

\subsubsection{Inference Interface Detection} \label{subsubsec:mapping-detection}
We analyze each parsed source file using a custom AST visitor that examines imports, assignments, function calls, and method calls.
For each inference backend, we define the inference interfaces targeted by the analysis based on the scope established in Section~\ref{subsec:inference-configuration-scope}. 
These backend-specific definitions identify the inference calls and configuration objects the analyzer recognizes, along with their associated configuration options.
During AST traversal, the analyzer determines whether encountered calls or configuration objects match one of these targeted interfaces.
A repository is considered to use a given inference backend when it detects at least one targeted interface in its source code. 

\subsubsection{Option Extraction \& Value Resolution}
When the analyzer detects a targeted inference interface, it examines the associated configuration to identify options explicitly specified by the developer and tries to resolve their values.

\paragraph{\textbf{Option Extraction.}}
An option is considered explicitly configured only when its name is explicitly specified in the repository together with a value expression and can be linked to a targeted inference interface.
Backend defaults and unnamed positional arguments are therefore not counted.
However, when a developer explicitly specifies a value equal to the backend default, the option is still considered explicitly configured.
The extraction includes, for example, direct keyword arguments such as \texttt{temperature=0.7}, configuration fields such as \texttt{config.temperature = 0.7}, and dictionary entries such as \texttt{\{"temperature": 0.7\}}. 
Options may also originate from command-line arguments or external configuration files when they can be statically linked to the targeted inference interface.

\paragraph{\textbf{Value Resolution.}}
For each extracted option, the analyzer attempts to resolve its value using several static-resolution strategies. 
These include following values propagated through variables, dictionaries, configuration objects, command-line parsers, and external configuration files.
When a command-line argument is used without being overridden in the repository, its statically defined default value is retained.
When an option can take multiple statically observable values, for example across conditional branches, we retain all observed alternatives.
For each extracted option, the analyzer records the corresponding inference backend, repository, file location, targeted inference interface, option name, resolved value when available, and the static-resolution strategy used.

\paragraph{\textbf{Unresolved Values.}}
Not all explicitly configured values can be resolved through static analysis.
Some depend on runtime inputs, environment variables, user-provided command-line arguments, dynamically constructed objects, external files that cannot be linked to the inference interface, or project-specific wrappers that cannot be resolved with sufficient confidence.
In such cases, the option is retained, and its value is recorded as \texttt{unresolved} rather than discarding the observation.
The resulting resolved and unresolved observations constitute the extracted configuration dataset.

\subsection{Iterative Refinement}\label{subsec:iterative-refinement}
The extraction process includes an iterative refinement loop. 
After an initial analysis pass, we inspect extraction errors, unresolved values, and recurring missed patterns to identify the analyzer's limitations. 
When a recurring pattern can be handled statically, we update the analyzer and re-analyze only the previously unresolved cases. 
This lets us iteratively improve value resolution without repeatedly processing the full dataset. 
Option values that still cannot be resolved after the run are kept as \texttt{unresolved}. 
After the final refinement, we re-ran the full process from the beginning to ensure no regressions and to produce the final dataset.

\subsection{Validation}

Given the scale of our study, exhaustive manual validation is impractical. We therefore use three complementary samples for each inference backend: $60$ detected configuration instances per backend ($180$ configurations in total), $30$ rejected repository candidates per backend ($90$ repositories in total), and $30$ unresolved option occurrences per backend ($90$ occurrences in total). The $60$ detected configurations are used to validate interface detection, option extraction, and resolved values. We choose this sample size to provide roughly a $95\%$ chance of observing at least one error if the error rate were $5\%$. The rejected candidates are used to check for missed interfaces, while the unresolved occurrences are used to understand why some values cannot be resolved.

For the $180$ detected configurations, interface-detection precision is $91.7\%$ for Transformers, $100\%$ for \gls{vllm}, and $98.3\%$ for OpenAI, corresponding to $96.7\%$ across the validation sample. The five Transformers false positives come from \texttt{generate} calls that were associated with Transformers although the receiver belonged to another backend or library. The single OpenAI false positive similarly involved an OpenAI-compatible call from a different library. Among the $90$ rejected repository candidates, we identify no missed target inference interface.

We then assess \emph{option extraction} on $163$ configurations, excluding six incorrect interface detections and $11$ cases for which dynamic argument propagation prevented a complete option-level ground truth. Within this manually validated subset, option-extraction precision is $100\%$ and recall is $98.9\%$, with an exact option-set match for $158/163$ configurations ($96.9\%$). The few missed options mainly result from indirect data flows through dictionaries or configuration objects.

Finally, we assess \emph{value resolution}. Among $208$ values marked as resolved and independently verifiable from the source code, we observe no errors. We also inspect the $90$ unresolved option occurrences, $30$ per backend, to check whether leaving them unresolved was justified. In $88$ cases, it was; the remaining two were simple literal values that could have been resolved statically.

\subsection{\textbf{Data Analysis}}
\label{subsec:data-analysis}

We exclude technical/internal options that do not affect inference or generation behavior, such as \texttt{trust\_remote\_code}. 
When we detect inference settings but cannot recover individual option names (e.g., \texttt{\_\_unexpanded\_kwargs\_\_}), we still consider the repository configured, but exclude these observations from RQ1--RQ3.

\paragraph{\textbf{RQ1: Inference-option usage.}}
We analyze each inference backend separately. For each option, we report its number of occurrences and repository-level adoption, and compare adoption across the categories defined in our study. For configured values, we consider only resolved observations and normalize equivalent representations, e.g., \texttt{0} and \texttt{0.0}. To limit the influence of repeated uses within a repository, each distinct value of an option is counted at most once per repository. We characterize value distributions using the most frequent values, the number of distinct values, and normalized Shannon entropy.

\paragraph{\textbf{RQ2: Inference Configuration Structure.}}
For each repository and inference backend, we compute the median number of distinct options across its configuration instances. 
We compare these distributions using pairwise two-sided Mann--Whitney $U$ tests with Holm correction and report Cliff's $\delta$ as effect size. 
We then focus on multi-option configurations and analyze option signatures at the repository level, counting each signature at most once per repository. 
We characterize diversity through the proportion of signatures observed in only one repository and concentration through top-$k$ repository coverage. 
We summarize concentration using AUC@100, the normalized area under the top-$k$ coverage curve for $k=1,\ldots,100$; higher values indicate that a smaller set of frequent signatures covers a larger share of repositories. 
Finally, we report the most frequent signatures by repository count together with their number of configuration instances.

\paragraph{\textbf{RQ3: Usage contexts.}}
We use GitHub repository topics as indicators of application context. Since topics are not standardized, we normalize them and manually group related labels into broader contexts derived from the most frequent topics in our dataset, while excluding labels unrelated to application domain or purpose. The second author reviewed the resulting contexts and topic mappings, and we discussed ambiguous cases until we agreed. 
We assign a repository to a context when at least one of its normalized topics maps to that context; repositories may belong to multiple contexts.

For each backend, context, and option, we compare option adoption at the repository level between repositories assigned to the context and those not assigned to it. 
Adoption is binary: we count a repository once if it configures the option at least once. We quantify the difference as
\[
\Delta_{pp} = P(o \mid c,b) - P(o \mid \neg c,b),
\]
where positive and negative values indicate higher and lower adoption in context $c$, respectively. 
We assess significance using two-sided Fisher's exact tests~\cite{fisher1922interpretation} and apply Benjamini--Hochberg correction~\cite{benjamini1995controlling} within each backend to control the false discovery rate. 
For presentation, we retain differences with $|\Delta_{pp}| \geq 5$ percentage points and adjusted $p \leq 0.05$; all comparisons are available in the replication package.
\section{Results}
\subsection{RQ1: Inference-option usage.}
Starting from the $10,283$ candidate repositories obtained after the repository discovery and selection stage (Section~\ref{subsec:repository-collection-selection}), our static analysis identified $5,398$ repositories using at least one targeted inference backend (Section~\ref{subsubsec:mapping-detection}). 
Backend usage is not mutually exclusive: $64.9\%$ of these repositories use a single backend, while $35.1\%$ use multiple backends. 
More specifically, $24.2\%$ use two backends and $11.0\%$ use all three. 
Among repositories using exactly two backends, the most frequent combination is Transformers and OpenAI, observed in $16.7\%$ of all validated repositories, followed by Transformers and \gls{vllm} ($6.9\%$) and \gls{vllm} and OpenAI ($0.5\%$).

Among the $5,398$ validated repositories, $5,118$ ($94.8\%$) contain evidence of at least one explicit inference configuration, while $280$ ($5.2\%$) contain none detected by our analysis. Across the option occurrences retained for RQ1--RQ3, we analyze $345,117$ occurrences.
Table~\ref{tab:rq1-overview} summarizes these observations by inference backend.

\begin{table}[t]
    \centering
    \scriptsize
    \caption{Overview of repositories and extracted inference option occurrences by inference backend.}
    \label{tab:rq1-overview}
    \begin{tabular}{lrrr}
        \toprule
        Backend    & Validated repos & Configured repos & Option occurrences \\
        \midrule
        \textbf{Transformers} & 3,000 & 2,505 & 102,107 \\
        \textbf{vLLM}         & 1,022 & 1,017 & 109,500 \\
        \textbf{OpenAI}       & 3,864 & 3,829 & 133,510 \\
        \midrule
        All                   & 5,398 & 5,118 & 345,117 \\
        \bottomrule
    \end{tabular}
\end{table}

\subsubsection{RQ1.1: Which \gls{llm} inference options do developers explicitly configure in practice?}

\begin{figure}[t]
   \centering
   \includegraphics[width=0.45\linewidth]{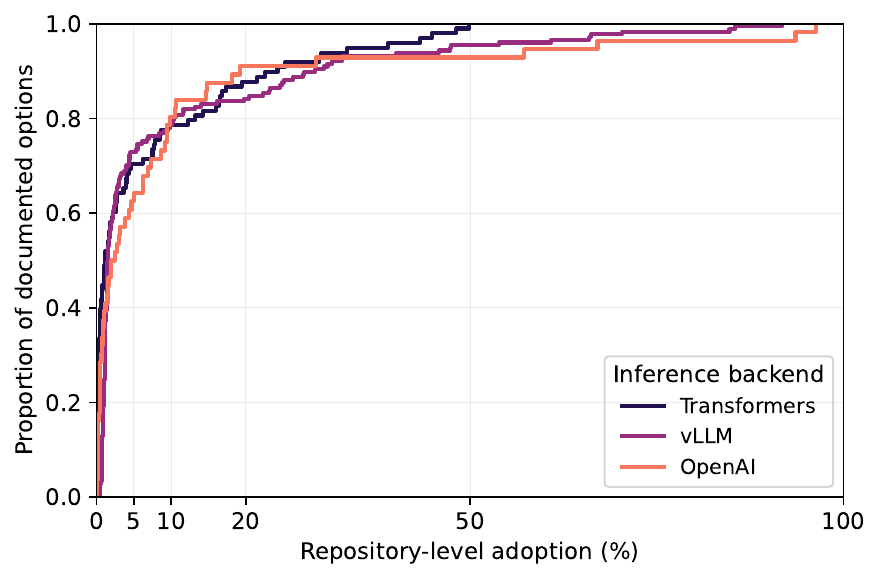}
   \caption{Cumulative distribution of repository-level adoption across documented options for each backend.}
\label{fig:rq1_documented_option_adoption_ecdf}
\end{figure}

\paragraph{\textbf{Adoption of the documented option space.}}
The extracted configurations cover almost all documented inference options considered in our study. 
We observe $95$ of the $98$ Transformers options ($96.9\%$), $176$ of the $177$ \gls{vllm} options ($99.4\%$), and all $56$ OpenAI options ($100\%$). 
Overall, $327$ of the $331$ documented backend-specific options ($98.8$\%) are observed at least once.
The only options never explicitly configured are \texttt{binary\_output}, \texttt{from\_flax}, and \texttt{video\_processor} for Transformers, and \texttt{chat\_template} for \gls{vllm}. 
The replication package includes full option coverage. 
However, this broad coverage does not imply common use. 
As shown in Figure~\ref{fig:rq1_documented_option_adoption_ecdf}, about $80\%$ of documented options across all three backends are configured in at most $10\%$ of repositories, and roughly $90\%$ are configured in at most $20\%$ of repositories. 
This indicates that most documented options are only rarely configured in practice.

\begin{figure}[t]
   \centering
   \includegraphics[width=\linewidth]{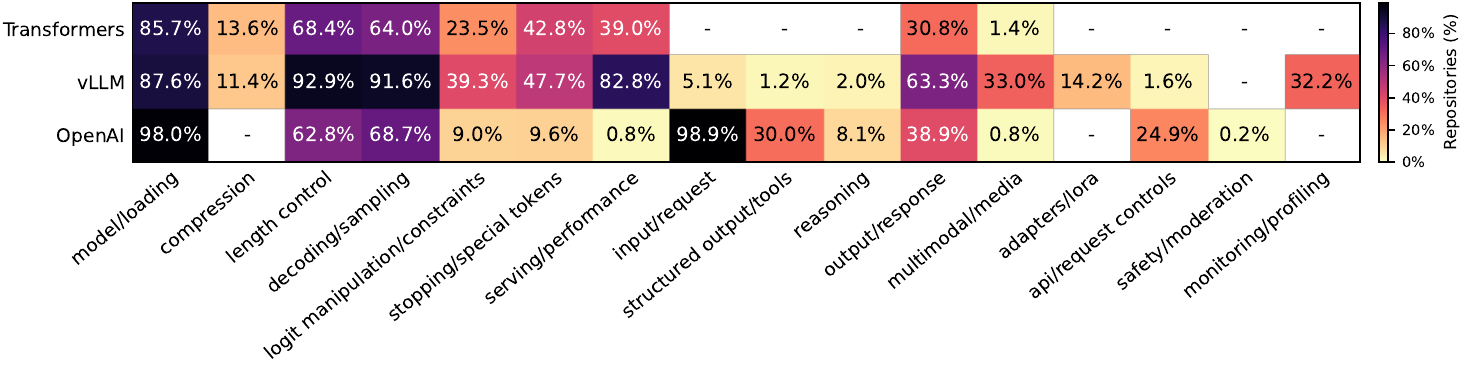}
   \caption{Most configured option categories by inference backend. -- indicates that no option from the corresponding category was observed for that backend.}
\label{fig:rq1-option-category-heatmap}
\end{figure}

\paragraph{\textbf{Configured option categories and frequent options.}}
\autoref{fig:rq1-option-category-heatmap} shows a common configuration core across the three inference backends. 
Model/loading, length control, and decoding/sampling are among the most frequent categories in all three. 
Model/loading options occur in $85.7\%$ of Transformers repositories, $87.6\%$ of \gls{vllm} repositories, and $98.0\%$ of OpenAI repositories. Length control and decoding/sampling are especially frequent in \gls{vllm}, where both appear in more than $90\%$ of repositories. 
OpenAI also shows very frequent use of input/request options ($98.9\%$).

\begin{figure}[t]
   \centering
   \includegraphics[width=\linewidth]{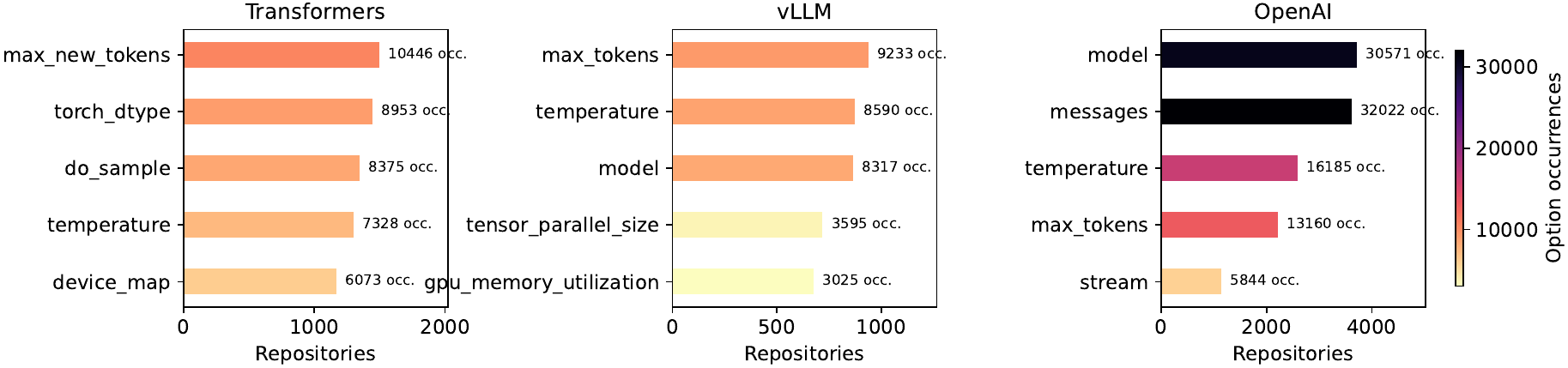}
   \caption{Most frequently configured inference options by repository count by inference backend, with occurrence counts shown as labels.}
\label{fig:rq1-top-options-by-framework}
\end{figure}

\autoref{fig:rq1-top-options-by-framework} shows that this common core is expressed through different options across backends. 
In Transformers, the most frequent options mix generation and model-loading settings, including \texttt{max\_new\_tokens}, \texttt{torch\_dtype}, \texttt{do\_sample}, \texttt{temperature}, and \texttt{device\_map}. 
In \gls{vllm}, generation options such as \texttt{max\_tokens}, \texttt{temperature}, and \texttt{top\_p} appear alongside serving-related options such as \texttt{tensor\_parallel\_size}, \texttt{gpu\_memory\_utilization}, and \texttt{max\_model\_len}. 
OpenAI is dominated by request-level options such as \texttt{model} and \texttt{messages}, together with generation controls such as \texttt{temperature} and \texttt{max\_tokens}.

The occurrence counts in \autoref{fig:rq1-top-options-by-framework} show that widely used options are also often configured many times. 
In Transformers, \texttt{max\_new\_tokens} accounts for $10,446$ occurrences. 
In \gls{vllm}, \texttt{max\_tokens} and \texttt{temperature} account for $9,233$ and $8,590$ occurrences. 
In OpenAI, \texttt{messages} and \texttt{model} stand out with $32,022$ and $30,571$ occurrences.

Beyond this shared core, some categories are more specific to individual backends. 
Serving/performance options occur in $82.8\%$ of \gls{vllm} repositories. 
OpenAI also shows frequent structured-output/tool configuration ($30.0\%$) and API/request controls ($24.9\%$). 
\gls{vllm} additionally shows frequent use of output/response ($63.3\%$) and multimodal/media ($33.0\%$) options.

The prominence of \texttt{model} also differs across backends because model selection is exposed differently. 
In \gls{vllm} and OpenAI, it is directly represented by the \texttt{model} option. 
In Transformers, model selection is spread across interfaces such as the optional \texttt{model} argument of \texttt{pipeline} and \texttt{pretrained\_model\_name\_or\_path} in \texttt{from\_pretrained}.

\subsubsection{RQ1.2: Which values do developers assign to these inference options?}
\paragraph{\textbf{Value resolution.}}
Overall, values are statically resolved for $266,173$ of the $345,117$ option occurrences retained in our analyses ($77.1\%$), while $78,944$ occurrences ($22.9\%$) remain unresolved. 
Resolution rates are $75.8\%$ for Transformers ($77,440$ of $102,107$ occurrences), $81.1\%$ for \gls{vllm} ($88,841$ of $109,500$), and $74.8\%$ for OpenAI ($99,892$ of $133,510$). 
The following analyses are based on these resolved occurrences.

\textbf{\textit{Value concentration and diversity.}}
Table~\ref{tab:rq1-value-summary} reports the three most common resolved values for the five options with the most value observations in each backend, together with the number of distinct values and their normalized Shannon entropy. Higher entropy indicates a more diverse value distribution. The complete distributions are available in the replication package.

For Transformers, value distributions vary substantially across options. \texttt{device\_map} is relatively concentrated ($H=0.49$), with \texttt{auto} accounting for $46.2\%$ of observations. Similarly, \texttt{torch\_dtype} is concentrated around \texttt{bfloat16} ($35.2\%$) and \texttt{float16} ($29.3\%$). In contrast, \texttt{max\_new\_tokens} is more diverse ($H=0.72$), spanning $121$ distinct values. \texttt{temperature} also shows no strongly dominant value, with \texttt{1}, \texttt{0}, and \texttt{0.7} accounting for $21.0\%$, $20.0\%$, and $16.4\%$ of observations, respectively.

For \gls{vllm}, model selection is particularly diverse, with $603$ distinct values and the highest entropy among the reported options ($H=0.90$); no individual model accounts for more than $1.6\%$ of observations. Output-length settings are also diverse, with \texttt{max\_tokens} reaching $H=0.73$ and \texttt{max\_model\_len} $H=0.78$. \texttt{gpu\_memory\_utilization} similarly spans a range of values ($H=0.75$), whereas \texttt{temperature=0} is a more recurring choice, accounting for $32.9\%$ of observations.

OpenAI shows a similar contrast. Model and output-length settings are relatively diverse: \texttt{model} spans $1,259$ distinct values ($H=0.76$), while \texttt{max\_tokens} reaches $H=0.74$. In contrast, \texttt{stream=true} accounts for $63.2\%$ of observations, and \texttt{top\_p=1} for $40.4\%$. For \texttt{temperature}, the most frequent value is \texttt{0} ($27.5\%$), followed by \texttt{0.7} ($17.4\%$) and \texttt{1} ($9.7\%$).

Overall, recurring decoding values are visible across backends, for example \texttt{top\_p=1}. Model and output-length settings tend to exhibit greater value diversity, whereas some control and execution-related options are more concentrated around recurring values.

\begin{table}[t]
    \centering
    \scriptsize
    \setlength{\tabcolsep}{2pt}
    \caption{Most common resolved values and diversity metrics for the five options with the most value observations in each inference backend. Each distinct value is counted at most once per repository.}
    \label{tab:rq1-value-summary}
    \resizebox{0.98\columnwidth}{!}{%
        \begin{tabular}{lrrrl}
            \toprule
            Option & Value observations & Entropy & \# Values & Top values \\
            \midrule
            \multicolumn{5}{l}{\textbf{Transformers}} \\
            \texttt{max\_new\_tokens} & 2,824 & 0.72 & 121 & \texttt{512} (12.6\%), \texttt{1024} (9.6\%), \texttt{128} (7.9\%) \\
            \texttt{torch\_dtype} & 2,386 & 0.62 & 11 & \texttt{bfloat16} (35.2\%), \texttt{float16} (29.3\%), \texttt{auto} (16.6\%) \\
            \texttt{temperature} & 1,801 & 0.66 & 35 & \texttt{1} (21.0\%), \texttt{0} (20.0\%), \texttt{0.7} (16.4\%) \\
            \texttt{do\_sample} & 1,720 & 0.64 & 3 & \texttt{true} (50.5\%), \texttt{false} (49.3\%), \texttt{null} (0.2\%) \\
            \texttt{device\_map} & 1,683 & 0.49 & 44 & \texttt{auto} (46.2\%), \texttt{cpu} (15.4\%), \texttt{cuda} (13.5\%) \\
            \midrule
            \multicolumn{5}{l}{\textbf{\gls{vllm}}} \\
            \texttt{model} & 2,746 & 0.90 & 603 & \texttt{Qwen2-7B-Instruct} (1.6\%), \texttt{opt-125m} (1.1\%), \texttt{Mistral-7B-Instruct-v0.3} (1.0\%) \\
            \texttt{max\_tokens} & 1,974 & 0.73 & 142 & \texttt{1} (12.4\%), \texttt{2048} (7.1\%), \texttt{512} (7.0\%) \\
            \texttt{temperature} & 1,335 & 0.66 & 34 & \texttt{0} (32.9\%), \texttt{1} (15.0\%), \texttt{0.7} (9.8\%) \\
            \texttt{max\_model\_len} & 1,062 & 0.78 & 92 & \texttt{4096} (9.1\%), \texttt{8192} (8.9\%), \texttt{32768} (7.6\%) \\
            \texttt{gpu\_memory\_utilization} & 916 & 0.75 & 32 & \texttt{0.9} (21.9\%), \texttt{0.8} (15.0\%), \texttt{0.5} (11.6\%) \\
            \midrule
            \multicolumn{5}{l}{\textbf{OpenAI}} \\
            \texttt{model} & 5,440 & 0.76 & 1259 & \texttt{gpt-4o} (9.7\%), \texttt{gpt-4o-mini} (7.9\%), \texttt{gpt-3.5-turbo} (6.0\%) \\
            \texttt{max\_tokens} & 4,110 & 0.74 & 147 & \texttt{4096} (10.1\%), \texttt{1024} (8.2\%), \texttt{2048} (5.9\%) \\
            \texttt{temperature} & 3,751 & 0.61 & 47 & \texttt{0} (27.5\%), \texttt{0.7} (17.4\%), \texttt{1} (9.7\%) \\
            \texttt{stream} & 1,368 & 0.61 & 3 & \texttt{true} (63.2\%), \texttt{false} (36.5\%), \texttt{null} (0.3\%) \\
            \texttt{top\_p} & 748 & 0.61 & 21 & \texttt{1} (40.4\%), \texttt{0.95} (19.7\%), \texttt{0.9} (15.6\%) \\
            \bottomrule
        \end{tabular}
    }
\end{table}

\begin{result}
\textbf{Answer to RQ1:}
Across the repositories in our dataset, $98.8\%$ of the documented inference options considered in our study are configured at least once; however, about $80\%$ are configured in at most $10\%$ of repositories across all three backends. 
Model/loading, length-control, and decoding/sampling options form a common core across backends, alongside backend-specific patterns such as serving/performance in \gls{vllm} and input/request in OpenAI. 
At the value level, recurring choices are visible across backends, for example \texttt{top\_p=1}, while model and output-length settings generally exhibit greater diversity.
\end{result}
\subsection{RQ2: Inference Configuration Structure.}
\begin{figure*}[t]
    \centering
    \begin{minipage}{0.48\textwidth}
        \centering
        \includegraphics[width=0.9\linewidth]{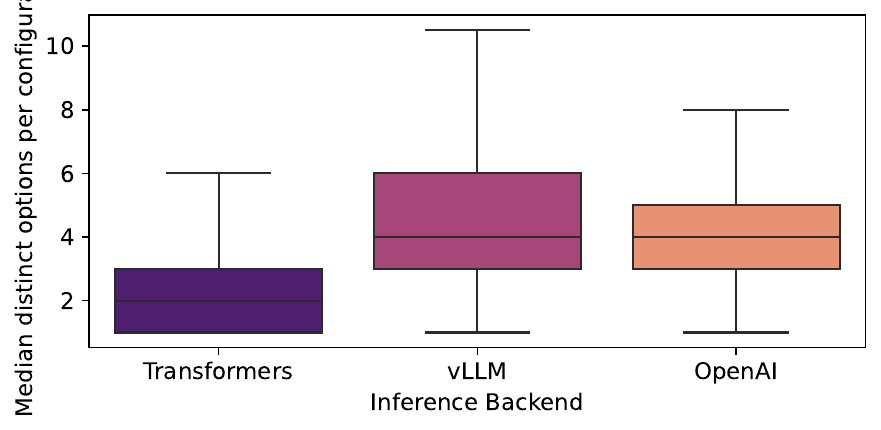}
        \captionof{figure}{Distribution of repository-level median configuration sizes.}
        \label{fig:rq2_repository_median_option_count_by_framework}
    \end{minipage}
    \hfill
    \begin{minipage}{0.48\textwidth}
        \centering
        \includegraphics[width=0.9\linewidth]{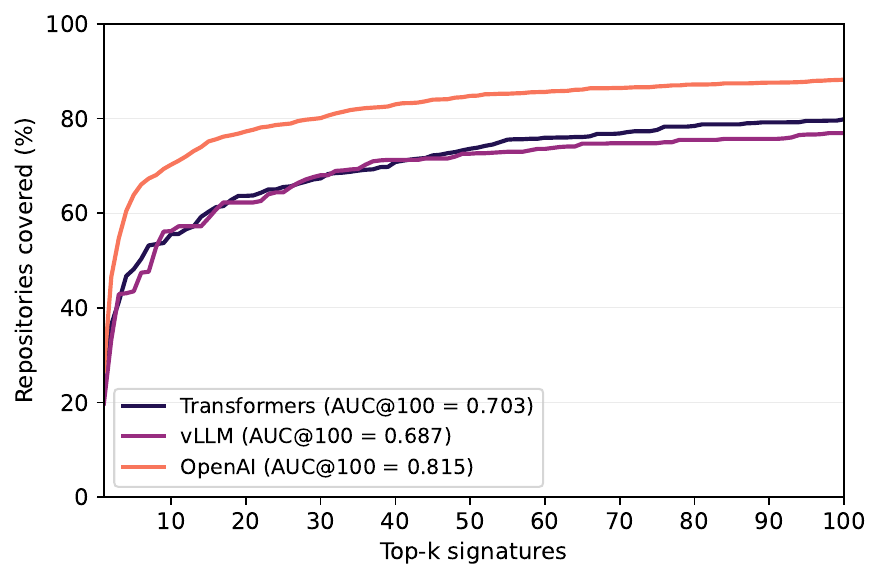}
        \captionof{figure}{Repository coverage by the top-$k$ multi-option signatures.}
        \label{fig:rq2-signature-top-k-coverage-by-framework}
    \end{minipage}

\end{figure*}

\textbf{\textit{Number of options per configuration.}}
Figure~\ref{fig:rq2_repository_median_option_count_by_framework} shows the distribution of repository-level median configuration sizes across inference backends. 
For each repository, we compute the median number of distinct options across its configuration instances. 
Transformers exhibits the smallest typical configuration size, with a median of $2$ options per repository, compared with $4$ for both \gls{vllm} and OpenAI.

Pairwise Mann--Whitney $U$ tests with Holm correction show significant differences between all three backends ($p_{\mathrm{adj}} < 0.001$).
The differences between Transformers and the other two backends are large, with Cliff's $\delta=-0.680$ for Transformers versus \gls{vllm} and $\delta=-0.632$ for Transformers versus OpenAI. Although \gls{vllm} and OpenAI have the same median, their distributions also differ significantly, with a smaller effect size ($\delta=0.169$).

Overall, configurations remain relatively small compared with the broad range of options observed in RQ1. 
Transformer configurations are typically smaller, while \gls{vllm} and OpenAI tend to include more options.

\textbf{\textit{Diversity and concentration of option signatures.}}
Among repositories containing at least one multi-option configuration, Transformers exhibits $1,652$ distinct signatures across $2,140$ repositories, with $61.6\%$ observed in only one repository. 
For \gls{vllm}, we observe $1,786$ signatures across $991$ such repositories, of which $56.8\%$ occur in only one repository. 
OpenAI exhibits $1,506$ signatures across $3,753$ such repositories, with $59.6\%$ observed in only one repository.

Figure~\ref{fig:rq2-signature-top-k-coverage-by-framework} shows the repository coverage achieved by the top-$k$ most frequent signatures. 
OpenAI shows the strongest concentration, with an AUC@100 of $0.815$. 
Transformers and \gls{vllm} show similar levels of concentration, with AUC@100 values of $0.703$ and $0.687$, respectively. 
This means the most frequent OpenAI signatures cover a larger share of repositories than those of the other two backends.

Overall, all three backends exhibit high signature diversity, with more than half of their signatures appearing in only one repository. 
OpenAI is more concentrated around its most frequent signatures, while Transformers and \gls{vllm} show similar concentration levels.

\begin{figure}[t]
   \centering
   \includegraphics[width=0.9\linewidth]{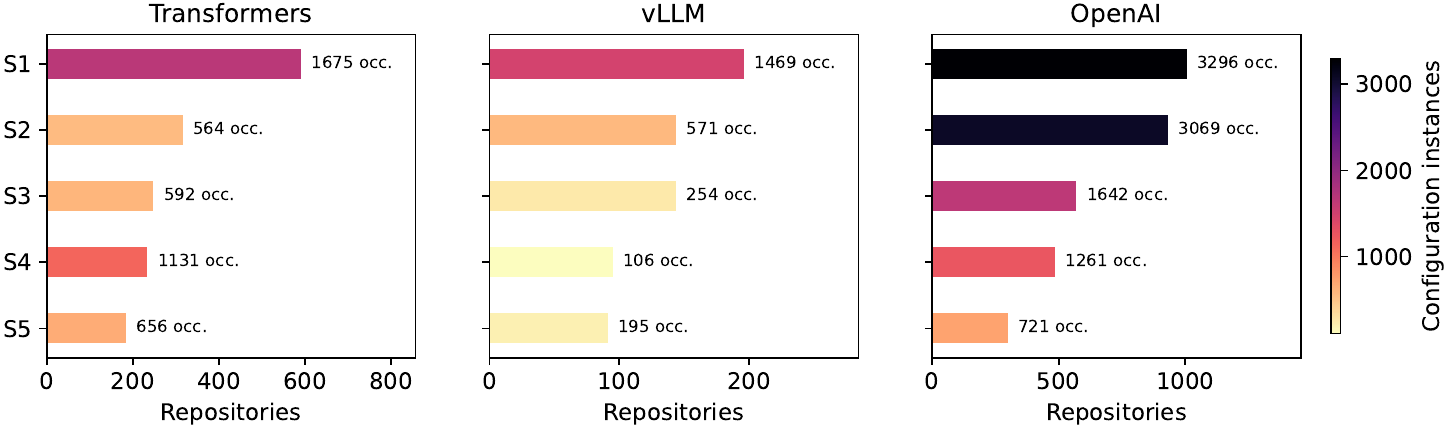}
   \caption{Most frequent multi-option signatures by repository count by inference backend, with configuration-instance counts shown as labels. Signature mappings are provided in the replication package.}
    \label{fig:rq2-top-option-signatures-by-framework}
\end{figure}

\textbf{\textit{Frequent option signatures.}}
Figure~\ref{fig:rq2-top-option-signatures-by-framework} shows the five most frequent multi-option signatures for each inference backend, ranked by repository count. 
The labels also report the number of configuration instances. 
The replication package includes the complete ranking.

For Transformers, the most prevalent signatures mainly involve model-loading options. 
The most frequent, \texttt{device\_map + torch\_dtype}, appears in $591$ repositories ($27.6\%$) and $1,675$ configuration instances. 
Several other frequent signatures also combine model-loading options, while generation-oriented combinations appear among the top signatures, notably \texttt{do\_sample + max\_new\_tokens}.

For \gls{vllm}, the most prevalent signature is \texttt{max\_tokens + temperature}, observed in $196$ repositories ($19.8\%$) and $1,469$ configuration instances. 
Another frequent signature adds \texttt{top\_p} to this combination, while the top five also contain larger signatures involving different sets of options.

For OpenAI, the most prevalent signatures share a recurring core around \texttt{messages} and \texttt{model}. 
The combination \texttt{messages + model} appears in $1,005$ repositories ($26.8\%$) and $3,296$ configuration instances, followed closely by \texttt{max\_tokens + messages + model + temperature} in $932$ repositories ($24.8\%$) and $3,069$ instances. 
Variants adding \texttt{temperature} or \texttt{max\_tokens} also occur among the most prevalent signatures.

Overall, the most frequent signatures differ across inference backends: Transformers frequently combine model-loading options, \gls{vllm} includes both small generation signatures and larger combinations, while OpenAI mainly combines request-level and generation options. These differences partly reflect the distinct configuration interfaces and abstraction levels exposed by each backend.

\begin{result}
\textbf{Answer to RQ2:}
Relative to their respective backend interfaces, inference configurations remain small compared with the available configuration spaces, with repository-level median sizes of $2$ options for Transformers and $4$ for vLLM and OpenAI. 
At the same time, configuration signatures are highly diverse, with more than half observed in only one repository. 
OpenAI configurations are more concentrated around their most frequent signatures, while Transformers and \gls{vllm} show similar concentration levels. 
The most frequent signatures also differ across backends, mainly combining model-loading options in Transformers, generation and larger option sets in \gls{vllm}, and request-level and generation options in OpenAI.
\end{result}

\subsection{RQ3: Application contexts.}

%\cq{This research question is exploratory: rather than attempting to construct an exhaustive application taxonomy, we use available GitHub topics to examine whether explicit configuration practices are associated with broad application contexts.}
%We identify application contexts from repository GitHub topics. 
To explore whether configuration practices vary across broad application contexts, we rely on repository GitHub topics. 
Among the $5,118$ repositories with at least one inference configuration, $2,287$ ($44.7\%$) provide at least one GitHub topic, comprising $6,029$ distinct topic labels. 
After normalization, $934$ topic labels map to at least one of our ten application contexts. 
Overall, $1,850$ repositories are assigned to one or more contexts, representing $36.1\%$ of all configured repositories. 
Among the remaining repositories, $2,831$ ($55.3\%$) provide no GitHub topics, while $437$ ($8.5\%$) provide topics that do not map to any of the retained contexts. 
The following analyses therefore focus on the $1,850$ mapped repositories.

\textbf{\textit{Distribution of usage contexts.}}
Table~\ref{tab:rq3-context-distribution} shows that inference configurations occur across all ten usage contexts, although their distribution is uneven. 
Agents and tool use is the most frequent context, associated with $49.4\%$ of mapped repositories, followed by multimodal and vision ($24.5\%$) and retrieval and knowledge systems ($19.8\%$). 
Serving and deployment accounts for $16.9\%$ of mapped repositories, followed by text and document processing ($13.8\%$) and code and software engineering ($13.2\%$). 
The remaining four contexts each occur in $10.0\%$ or fewer of mapped repositories.

\begin{table}[t]
    \centering
    \scriptsize
    \caption{Distribution of repositories across usage contexts. Percentages are computed over the repositories mapped to at least one context. A repository may belong to multiple contexts.}
    \label{tab:rq3-context-distribution}
    \begin{tabularx}{\columnwidth}{Xrr}
        \toprule
        \textbf{Usage context} & \textbf{Repositories} & \textbf{Percentage} \\
        \midrule
        Agents \& tool use             & 914 & 49.4\% \\
        Multimodal \& vision           & 454 & 24.5\% \\
        Retrieval \& knowledge systems & 366 & 19.8\% \\
        Serving \& deployment          & 312 & 16.9\% \\
        Text \& document processing    & 255 & 13.8\% \\
        Code \& software engineering   & 244 & 13.2\% \\
        Speech \& audio                & 185 & 10.0\% \\
        Evaluation \& benchmarking     & 167 & 9.0\% \\
        Conversational assistants      & 158 & 8.5\% \\
        Security \& safety             & 154 & 8.3\% \\
        \bottomrule
    \end{tabularx}
\end{table}

\textbf{\textit{Context-specific configuration practices.}}
Table~\ref{tab:rq3-characteristic-options} reports the option--context combinations that meet the statistical criteria defined in Section~\ref{subsec:data-analysis}. 
Across backends, the retained differences concern a relatively small subset of options, although they span several application contexts.

For Transformers, the observed differences involve different types of configuration options depending on the context. 
Multimodal and vision repositories more frequently configure generation options such as \texttt{use\_cache}, \texttt{max\_new\_tokens}, and \texttt{num\_beams}, with differences of up to $+17.5$~pp. 
Speech and audio repositories more frequently configure \texttt{device}, \texttt{model}, \texttt{task}, \texttt{batch\_size}, and \texttt{feature\_extractor}, while \texttt{device\_map} and \texttt{use\_fast} are less frequent. 
Among these options, \texttt{feature\_extractor} is directly involved in preprocessing inputs such as audio signals. 
Evaluation and benchmarking repositories more frequently configure several memory, length, and stopping-related options, while serving and deployment repositories more frequently configure \texttt{dtype}. 
Retrieval and knowledge systems and text and document processing repositories instead show lower adoption of several generation and model-loading options.

For \gls{vllm}, context-specific differences are much more limited. 
The only retained difference concerns multimodal and vision repositories, where \texttt{limit\_mm\_per\_prompt} is substantially more frequent ($+30.7$~pp). 
This option controls the number of multimodal inputs allowed per prompt.

For OpenAI, differences are observed across several contexts. 
Agent-oriented repositories more frequently configure \texttt{tools} and \texttt{tool\_choice}, which control tool availability and selection during inference. 
Conversational assistants and serving and deployment repositories more frequently configure \texttt{stream}; serving repositories also more frequently configure \texttt{stream\_options}. 
These options control streamed response delivery. 
Evaluation and benchmarking repositories more frequently configure generation controls such as \texttt{max\_tokens} and \texttt{temperature}, while code and software engineering repositories show higher adoption of \texttt{reasoning\_effort}. 
In contrast, multimodal and vision repositories configure streaming and tool-related options less frequently, while text and document processing repositories also show lower adoption of \texttt{tools}.

\begin{table*}[t]
    \centering
    \scriptsize
    \caption{Context-specific differences in inference-option adoption. Positive and negative values indicate over and under representation, respectively; pp denotes percentage points}
    \label{tab:rq3-characteristic-options}
    \begin{tabularx}{\textwidth}{p{0.13\textwidth} p{0.25\textwidth} X}
        \toprule
        \textbf{Backend} & \textbf{Usage context} & \textbf{Characteristic options} \\
        \midrule
        Transformers & Code \& software engineering & \texttt{use\_fast} ($-19.6$\,pp) \\
        Transformers & Conversational assistants & \texttt{model} ($+20.1$\,pp) \\
        Transformers & Evaluation \& benchmarking & \texttt{low\_cpu\_mem\_usage} ($+19.4$\,pp), \texttt{min\_length} ($+11.8$\,pp), \texttt{stop\_strings} ($+9.0$\,pp), \texttt{bad\_words\_ids} ($+8.0$\,pp) \\
        Transformers & Multimodal \& vision & \texttt{max\_new\_tokens} ($+17.5$\,pp), \texttt{use\_cache} ($+16.8$\,pp), \texttt{num\_beams} ($+13.8$\,pp), \texttt{do\_sample} ($+12.6$\,pp), \texttt{low\_cpu\_mem\_usage} ($+9.1$\,pp), \texttt{local\_files\_only} ($+7.3$\,pp), \texttt{model} ($-10.4$\,pp) \\
        Transformers & Retrieval \& knowledge systems & \texttt{top\_p} ($-15.1$\,pp), \texttt{num\_beams} ($-12.5$\,pp), \texttt{use\_cache} ($-11.7$\,pp), \texttt{repetition\_penalty} ($-11.7$\,pp) \\
        Transformers & Serving \& deployment & \texttt{dtype} ($+17.1$\,pp), \texttt{local\_files\_only} ($+10.5$\,pp), \texttt{eos\_token\_id} ($-12.2$\,pp) \\
        Transformers & Speech \& audio & \texttt{device} ($+22.2$\,pp), \texttt{model} ($+19.4$\,pp), \texttt{task} ($+12.6$\,pp), \texttt{dtype} ($+12.4$\,pp), \texttt{feature\_extractor} ($+11.4$\,pp), \texttt{batch\_size} ($+11.3$\,pp), \texttt{use\_fast} ($-22.8$\,pp), \texttt{device\_map} ($-21.5$\,pp), \texttt{low\_cpu\_mem\_usage} ($-12.7$\,pp) \\
        Transformers & Text \& document processing & \texttt{max\_length} ($+12.5$\,pp), \texttt{tokenizer} ($+11.0$\,pp), \texttt{torch\_dtype} ($-15.3$\,pp), \texttt{attn\_implementation} ($-13.2$\,pp), \texttt{use\_cache} ($-12.3$\,pp) \\
        \midrule
        \gls{vllm} & Multimodal \& vision & \texttt{limit\_mm\_per\_prompt} ($+30.7$\,pp) \\
        \midrule
        OpenAI & Agents \& tool use & \texttt{tools} ($+14.3$\,pp), \texttt{stream} ($+8.1$\,pp), \texttt{tool\_choice} ($+7.7$\,pp), \texttt{input} ($+7.1$\,pp) \\
        OpenAI & Code \& software engineering & \texttt{reasoning\_effort} ($+6.5$\,pp) \\
        OpenAI & Conversational assistants & \texttt{stream} ($+25.2$\,pp), \texttt{stream\_options} ($+9.0$\,pp) \\
        OpenAI & Evaluation \& benchmarking & \texttt{max\_tokens} ($+18.1$\,pp), \texttt{temperature} ($+14.8$\,pp), \texttt{prompt} ($+9.1$\,pp) \\
        OpenAI & Multimodal \& vision & \texttt{stream} ($-12.0$\,pp), \texttt{tools} ($-9.0$\,pp) \\
        OpenAI & Serving \& deployment & \texttt{stream} ($+17.0$\,pp), \texttt{tools} ($+12.0$\,pp), \texttt{stream\_options} ($+10.7$\,pp) \\
        OpenAI & Text \& document processing & \texttt{tools} ($-13.2$\,pp) \\
        \bottomrule
    \end{tabularx}
\end{table*}

\begin{result}
\textbf{Answer to RQ3:}
Configuration practices differ across application contexts, but the differences are concentrated in a subset of options. 
Some of these differences involve options whose function directly matches the context, such as \texttt{feature\_extractor} in speech and audio repositories, \texttt{limit\_mm\_per\_prompt} in multimodal repositories, \texttt{tools} and \texttt{tool\_choice} in agent-oriented repositories, and streaming options in conversational and serving repositories. 
Other contexts differ instead in broader generation, model-loading, or execution options.
\end{result}
\section{Discussion}\label{sec:discussion}

\paragraph{\textbf{\textit{The Configuration Space Used in Practice}}}
Our results characterize the part of the configuration space that developers explicitly manipulate, rather than identifying which options matter most for inference. 
How often an option is configured, or how often a particular value is selected, should therefore not be interpreted as a measure of its importance or impact on inference behavior. 
Frequently configured options may simply be more visible because they are more commonly documented, discussed, or studied. 
Conversely, rarely configured options may still matter because developers can rely on defaults, may be unaware of the option or its effect, or may need it only in specific situations.

\textbf{\textit{Backend Roles}}
The differences in configuration structure observed across backends appear closely related to their respective roles and abstraction levels.
Transformers exposes decisions related to model loading, local execution, and generation, whereas \gls{vllm} additionally exposes serving concerns such as parallelism, memory utilization, batching, and execution. 
OpenAI, in contrast, abstracts most deployment-level decisions and primarily exposes request-level controls, including generation, output formatting, tool use, and streaming. 
These differences may partly explain the configuration-size differences observed in RQ2. 
Transformer configurations are typically smaller, with a median of $2$ options compared with $4$ for both \gls{vllm} and OpenAI. 
For \gls{vllm}, this larger configuration size can involve additional serving dimensions such as tensor parallelism, GPU memory utilization, or batching, while OpenAI reaches a similar median through different request-level controls.

The studied backends should also not necessarily be viewed as mutually exclusive alternatives. 
Their different roles can make them complementary within the same project, which is consistent with the overlap observed in our dataset: $35.1\%$ of validated repositories use more than one backend, including $24.2\%$ using two and $11\%$ using all three. 
The most common two-backend combination is Transformers with OpenAI, followed by Transformers with \gls{vllm}. 
This overlap suggests that backend choice may reflect different inference needs or abstraction levels within the same software system.

\textbf{\textit{Frequent Option Signatures}}
The recurring option signatures observed in RQ2 further show that configuration choices are often organized around groups of related options. 
Frequent signatures combine model-loading options in Transformers, generation and serving-related options in \gls{vllm}, and request-level and generation options in OpenAI. 
Such recurring combinations may reflect functional relationships between options, but they may also result from shared examples, documentation, reused code, or common configuration practices. 
Their frequency should therefore not be interpreted as evidence that these combinations are optimal or deliberately selected as a unit. 
Rather, they reveal recurring structures in how developers configure inference in practice.

\textbf{\textit{Application Context Influence}}
Application context measurably but selectively influences the inference options developers configure. 
Only a subset of options shows substantial and statistically significant differences across contexts. 
Some of these differences directly match the corresponding application needs, such as tool-related options in agent applications, streaming options in conversational and serving repositories, or multimodal-specific options in multimodal applications. 
For many other options, however, configuration practices remain relatively similar across contexts. 
One explanation is that many inference decisions are generic enough to be reused across application contexts. 
Another is that configuration choices may be driven more strongly by factors such as workload scale, hardware constraints, model characteristics, or deployment environment than by application context itself. 
The limited differences observed across contexts may also reflect limited context-specific adaptation: developers may reuse similar options even when different application contexts could benefit from different choices, particularly because it remains unclear whether the same options and values have comparable effects on metrics such as latency, energy consumption, or output quality across contexts.

\textbf{\textit{Implications for Practice and Research}}
Our findings have implications for developers, inference backend maintainers, and researchers. 

For developers, the configurations observed in real-world repositories provide descriptive reference points for understanding which parts of the inference configuration space other projects make explicit.
However, they should not be interpreted as recommended or optimal configurations. 
Common practices may reflect conventions, available examples, or existing guidance rather than deliberate optimization. 
They are therefore better viewed as examples of configurations used in practice, which developers can adapt to their own workloads.

For inference backend maintainers, our results show which parts of the configuration space developers commonly use and which remain less visible in practice. 
Beyond documenting individual options, guidance could more clearly connect groups of related options to concrete goals and usage conditions. 
For example, documentation could indicate which options are particularly relevant for throughput, memory usage, long-context inference, output control, or specific application settings. 
This could help developers understand not only what an option does, but also when it is worth configuring.

For researchers, our results can help make the large inference configuration space more tractable to study. 
Exhaustively evaluating all options, values, combinations, models, and application contexts is impractical. 
The options and combinations observed in practice can therefore help prioritize which parts of this space are most relevant to investigate. 
At the same time, rarely configured options should not be overlooked, as their limited adoption may reflect limited awareness rather than usefulness. 
More broadly, the patterns observed in our study highlight the importance of understanding not only the effects of configuration choices, but also the factors that drive how developers make these choices.
\section{Threats to Validity}
\label{sec:threats}

\paragraph{\textbf{Construct Validity}}
Our analysis relies on documented options from specific versions of Hugging Face Transformers, \gls{vllm}, and OpenAI, and may therefore miss undocumented, deprecated, renamed, or version-specific options. 
Static analysis may also fail to resolve values that depend on runtime inputs, environment variables, wrappers, or dynamically constructed objects. Overall, we resolve $266,173$ of $345,117$ option occurrences ($77.1\%$). 
Finally, repositories may contain several configurations across files or execution paths, so option occurrences should not be interpreted as unique project-level choices.

\textbf{Internal Validity}
Implementation errors may affect option extraction, resolution, or classification, particularly when handling aliases, wrappers, or \texttt{**kwargs}. 
Our analysis also focuses primarily on Python code and selected configuration formats; configurations defined in notebooks, shell scripts, deployment manifests, Docker files, YAML files, or environment variables may therefore be missed or only partially captured.

\textbf{External Validity}
We consider only public GitHub repositories and restrict our dataset to Python projects with recent activity and at least 50 stars. 
These repositories may not represent smaller, older, non-Python, private, or industrial projects~\cite{kalliamvakou2014promises}. 
Repository discovery also depends on GitHub search queries, metadata, and filtering criteria, which may miss projects whose backend usage is not easily searchable. 
Finally, we study only Hugging Face Transformers, \gls{vllm}, and OpenAI; other inference backends may exhibit different configuration practices.

\textbf{Conclusion Validity}
Our context analysis relies on GitHub topics, which may be incomplete, outdated, or overly general. 
Context-level findings should therefore be interpreted as trends rather than exhaustive classifications. 
Consider differences in repository counts across backends when interpreting cross-backend comparisons.
\section{Conclusion \& Future Work}\label{sec:discussion}
\gls{llm} inference configuration is becoming an increasingly important aspect of deploying \gls{llm}-based software, with growing research interest in understanding how configuration choices affect performance, energy consumption, and accuracy. 
In this paper, we complement these studies with a large-scale empirical analysis of the inference options developers explicitly configure in open-source \gls{llm} inference software.

Our results reveal a clear gap between the large configuration spaces exposed by inference backends and the much smaller set of options that developers use frequently. 
Across the studied repositories, nearly all documented options in scope appear at least once, yet approximately 80\% are explicitly configured in at most 10\% of repositories using the corresponding backend. 
Individual configurations are small relative to the available configuration spaces, while the combinations of options used across projects remain highly diverse and reflect the different abstractions each backend exposes. 
Our exploratory context analysis further identifies selected associations between application contexts and option adoption. 
These findings characterize configuration practice, not configuration effectiveness: frequently observed choices are not necessarily optimal, and rarely explicit options may remain important through backend defaults.
Future work can connect these observed practices with controlled measurements of latency, throughput, energy consumption, monetary cost, and output quality to determine whether the options developers frequently manipulate are also the most consequential and whether rarely configured options represent overlooked optimization opportunities.

\textbf{Data Availability Statement:}
All artifacts (code, datasets, and results) are made available in an online replication package~\cite{replicationpackage}.

\newpage
\bibliographystyle{ACM-Reference-Format}
\bibliography{bibfile.bib}
\end{document}